\documentclass[a4paper]{article}

\usepackage{amsmath,amsfonts,amssymb,amsthm}

\begin{document}

\title{On the degeneracy of the energy levels of Schr\"odinger and Klein-Gordon equations  on Riemannian coverings}

\author{Claudia Maria Chanu, Giovanni Rastelli\\ \\ Dipartimento di Scienze Umane e Sociali, \\ Universit\`{a} della Valle d'Aosta\\
		Dipartimento di Matematica, Universit\`{a} di Torino,\\
		email: c.chanu@univda.it; giovanni.rastelli@unito.it}
	
	\maketitle

\begin{abstract}
We study the degeneracy of the energy levels of the Schr\"odinger equation with Kepler-Coulomb potential and of the Klein-Gordon equation on Riemannian coverings of the Euclidean space and of the Schwarzschild space-time respectively. Degeneracy of energy levels is a consequence of the superintegrability of the system. We see how the degree of degeneracy changes depending on the covering parameter $k$, the parameter that in space-times can be related with a cosmic string, and show examples of lower degeneracy in correspondence of non integer values of $k$.\end{abstract}

\

\newtheorem{defi}{Definition}
\newtheorem{teo}{Theorem}
\newtheorem{con}{Conjecture}
\newtheorem{prop}[teo]{Proposition}
\newtheorem{cor}[teo]{Corollary}
\newtheorem{lem}[teo]{Lemma}
\newtheorem{rmk}{Remark }
\newtheorem{exm}{Example}

\renewcommand{\theenumi}{\roman{enumi}}
\renewcommand{\labelenumi}{\theenumi)}


 \section{Introduction}
In \cite{CR} we began to study the problem of separation of variables and integrability on Riemannian coverings for classical finite dimensional Hamiltonian systems. Now we consider some specific example, in order to enlighten  what happens in the quantum case, when we examine a quantum system defined on a Riemannian covering instead of on a ordinary Riemannian manifold.  

In the classical case the question of the globality arises since
in the literature \cite{TTW,PW,MPW,TTWcdr,CDRPW} natural Hamiltonians
of the form
 \begin{equation*}
H=\frac 12 p_r^2 +\frac 1{2r^2}\left( p_\phi^2  +V(\sin(k
\phi), \cos(k\phi))\right) +F(r)
 \end{equation*}
are considered as Hamiltonian in polar coordinates defined on the  Euclidean plane for any real non zero value of $k$.
In some of these papers \cite{TTWcdr,CDRPW} these Hamiltonians are
rewritten, by the rescaling $\Pi: \psi=k\phi$, as
\begin{equation}\label{ham}
 H=\frac 12 p_r^2 +\frac {k^2} {2 r^2}\left( p_\psi^2  +\frac{1}{k^2}V(\sin(\psi), \cos(\psi))\right) +F(r)   
\end{equation}
and interpreted as Hamiltonians on manifold with a warped product metric.  These Hamiltonians have been considered recently in several articles devoted to the study of the superintegrability of Hamiltonian systems, both in the classical and in the quantum case \cite{TTW,PW,MPW,TTWcdr,CDRPW}. 
Metrics of this form appear in the classical and quantum Hamiltonians of the Tremblay-Turbiner-Winternitz system. This system, derived from the Smorodinski-Winternitz system, attracted much interest in recent years because it admits non-trivial first integrals of high degree, or symmetry operators of high order in the quantum version, the degree/order depending on the parameter k. The same type of metrics, in four dimensional spacetimes, is related with the presence of cosmic strings of parameter k.

In \cite{CR} we start the study of  the global structure of these manifolds.
Indeed, we consider the metric  
of the rescaled Hamiltonian (\ref{ham})
as a metric on a Riemannian covering of the Euclidean plane and enlighten the differences of this alternative approach. The definition and some additional feature of Riemannian coverings are recalled in Section \ref{s:cover}. Thus, a class of Riemannian manifolds whose metric tensor depends on a integer, rational or even real parameter $k$, is considered. 

We remark that even if the potential of the natural Hamiltonian depends on $r$ only, as in the Kepler system, the  globality of the first integrals strongly depends on the value of $k$. As instance the superintegrability of Kepler system on a Riemannian covering depends on $k$: for non rational values of $k$ the system is not superintegrable, and for non integer rational values of $k$ the system is superintegrable but not more quadratically superintegrable. These results agree with  \cite{ AW,BKM}, where the superintegrability of the Kepler system on Riemannian coverings of the Euclidean plane with $k<1$, i.e. cones, is proved true if and only if $k$ is rational.

The main result of \cite{CR}  is about the superintegrability of the Tremblay-Turbiner-Winternitz, the Post-Winternitz and, consequently, of the harmonic oscillator and Kepler-Coulomb systems. The introduction of Riemannian coverings allows us to understand in a better way issues of global definition for the first integrals of high degree of those systems determined in \cite{TTW, PW}, and allow us to show the global definition of those determined via the extension procedure in \cite{TTWcdr,CDRPW}. In particular, in \cite{CR} it is shown that the Kepler-Coulomb system on Riemannian coverings does not admit as first integral the standard Laplace vector,  instead, it is replaced by a first integral of degree depending on $k$. Similar considerations about Laplace vector and superintegrability on cones can be found in \cite{ AW,BKM}.

The aim of our present work is to look at the changes in the notions of \textbf{degeneracy and superintegrability} in the quantum case, when this kind of Riemannian manifolds are involved. As in \cite{CR}, we consider the configuration manifolds as Riemannian coverings of the manifold corresponding to $k=1$: the Euclidean space or the Schwarzschild space-time in the present case. In this way, we see from a geometric view point that problems of global definition of the eigenfunctions of the Hamiltonian operator can arise for non integer values of $k$, even if the Hamiltonian operator itself is globally defined for any values of $k$, as it happens for the Kepler system. We see how Schr\"odinger equation for the Kepler-Coulomb potential on $\mathbb E^3$ and Klein-Gordon equation of  on Schwarzschild space-time change when they are mapped on the respective Riemannian coverings by the covering map $\Pi$, being $\phi$ the cyclic coordinate on $\mathbb S^2$ \cite{CR}.

We find that the degeneracy of the energy levels can decrease for certain non integer values of $k$, for both the systems we are considering, in accordance with the behaviour of the classical cases studied in \cite{CR}.

Section \ref{s:schr} is devoted to the quantum version of the 3-dimensional Kepler system: the Schr\"odinger equation for the hydrogen atom is studied when the configuration manifold is a Riemannian covering of the Euclidean space.

In section \ref{s:schw}, we consider as a second example  the Klein-Gordon equation  on the Riemannian coverings of the Schwarzschild four-dimensional space-time.

Riemannian manifolds depending on a parameter that can be considered as Riemannian coverings appear in recent studies  describing accelerated black holes and cosmic strings \cite{App,Vil}. Metrics of this type do appear also in  orbifold theory (cone manifolds and ``footballs"), see for example \cite{coop}, where it is made use of the discrete, dihedral symmetries introduced by  the parameter in the metric. Another application is made in \cite{DG} to represent circularly symmetric N-vortex solutions of elliptic Sinh-Gordon and Tzitzeica equation.  
 
\section{Coverings and superintegrability}\label{s:cover}
We recall the definition of Riemannian coverings as introduced in \cite{IC} and employed in \cite{CR}.  
\begin{defi} If $M$ and $N$ are connected topological manifolds, we say that a map $\Pi: M\to N$ is a covering if every $p \in N$ has a connected open neighborhood $U$ such that $\Pi$ maps each component of $\Pi^{-1}[U]$ homeomorphically onto $U$.

If $M$ and $N$ are also differentiable manifolds, then $\Pi$ is a differentiable covering if $\Pi $ is differentiable of maximum rank on $M$.

If $M$ and $N$ are also Riemannian manifolds, then $\Pi$ is a Riemannian covering if $\Pi $ is differentiable of maximum rank and a local isometry of $M$ on $N$ (i.e. the metric on $M$ is the pull-back via $\Pi$ of the metric on $N$ wherever the Jacobian linear map $\Pi_*:TM \to TN$ is one-to-one).

\end{defi}

In \cite{CR} we deal both on the coverings of the Euclidean plane and on the sphere $\mathbb S^2$. 
In the present paper we deal with metric that are warped products of the metric of the sphere with an additional manifold.

The metric of the Euclidean three-space in spherical coordinates is
\begin{equation}\label{m1}
    ds^2=dr^2
+r^2 \left(d\theta ^2+\sin(\theta)^2 d\phi^2\right),
\end{equation}
We  now consider the metric 
 \begin{equation*}
G_k=dr^2+ r^2\left(d\theta ^2+k^2 \sin^2 \theta \, d\phi^2\right),
 \end{equation*}
with $k\in \mathbb R^+$, $0<r$, $0\leq \theta < \pi$, $0\leq \phi < 2\pi$. $G_k$ is locally the metric of the Euclidean three-space $\mathbb E^3$ and for $k=1$ it globally coincides with it. By the rescaling $\Pi$,  we get
 \begin{equation*}
G= dr^2+ r^2 \left(d\theta^2+\sin^2\theta d\psi^2\right).
 \end{equation*}
Hence, we have again that each dihedral angle of width $2\pi/k$ in $\phi$  is mapped isometrically by $(\theta, \psi)$ in $\mathbb E^3$.

We have the same general non-globality of the coordinates whenever $k$ is not integer. The map \begin{equation*}
    (r, \theta,\psi) \mapsto (r,\theta ,k\phi)
\end{equation*}
 has the same structure of the map $(r,\phi)$ of the case of the plane shown in Section 3 of \cite{CR}.

\begin{prop}{\label{p:cover}}
    For $k$ integer and $k \geq 1$, the map $\Pi: \psi=k\phi$ is a
Riemannian covering map.
\end{prop}

\begin{rmk} \rm
	  Whenever $k$ is not integer, $M_k$ fails to be a covering of $M_1$ in the strict sense used in \cite{IC}, because not all the points in $M_k$ have neighborhoods that can be projected on $M_1$ consistently with its topology.   Indeed, in these cases the projections on $M_1$ of  neighborhoods of points with $\psi=0$ and $\psi=2k\pi$ cannot be made to coincide. For simplicity, however, we call coverings also the $M_k$ with $k$ not integer, and we leave to the nature of $k$ the distinction between the  coverings in a strict sense and those that are not such.
	 \end{rmk}

  When $k<1$, the Riemannian covering of $\mathbb S^2$ generated by $\psi$ is sometime called a {\it football}.

We recall that
\begin{defi} \rm
A $n$-dimensional Hamiltonian system is (maximally) \emph{superintegrable} when it admits $2n-1$ functionally independent and globally defined first integrals; if the first integrals are all quadratic in the momenta, the system is called \emph{quadratically superintegrable}.     
\end{defi} 

Superintegrability in the quantum case is defined in the same way, with the difference that we have quantum symmetry operators instead of first integrals, and polynomial independence, meaning that  no polynomial in the symmetry operators  formed entirely using Lie
anticommutators should vanish identically, instead of functional independence \cite{MPW}.

For classical systems, maximal superintegrability manifests itself in the closure of bounded orbits, for quantum ones, in the maximal degeneration of the energy levels \cite{MI}. The Kepler-Coulomb system in the plane, or in the three-dimensional Euclidean space, is maximally superintegrable both in its classical and quantum version. The maximal degeneration of the energy levels of the Kepler-Coulomb system is made explicit by the well known complete set of quantum numbers.

What is important for us, is that maximal degeneration of the energy levels is a consequence of the maximal  superintegrability of the quantum system. This fact was first remarked by Fock in 1935, see for example \cite{MI,MPW} and references therein. 

In the following, we analyze the energy levels   of the Schr\"odinger equation on Riemannian coverings of $\mathbb E^3$ and, consequently,  of the Klein-Gordon equation on Riemannian coverings of the Schwarzschild background, and show how their degeneration depends on the covering parameter $k$.

\section{Schr\"odinger equation for the hydrogen atom}\label{s:schr}

We recall that the Kepler-Coulomb quantum system is maximally quadratically superintegrable, meaning that it admits five independent symmetry operators all of order two.

Given the metric  
\begin{equation*}\label{m1old}
    ds^2=dr^2
+r^2 \left(d\theta ^2+\sin(\theta)^2 d\phi^2\right),
\end{equation*}
in standard spherical coordinates of $\mathbb E^3$, by the map $\Pi$, $k\in \mathbb R-\{0\}$, we can pull back (\ref{m1}) on a Riemannian covering of $\mathbb E^3$ obtaining the metric 
\begin{equation*}\label{m2}
    ds^2=dr^2
+r^2 \left(d\theta ^2+k^2\sin(\theta)^2 d\phi^2\right).
\end{equation*}
We now consider the time-independent Schr\"odinger equation on the covering,

\begin{equation}\label{SE}
    -\frac 12 \Delta \Psi+V\Psi=h\Psi,\quad h\in \mathbb R,
\end{equation}

with the assumption that the Plank constant satisfies $\hbar^2=1$ and where $\Delta$ is the Laplace-Beltrami operator 
\begin{equation*}
    \Delta \Psi=\frac 1{\sqrt{g}}\partial_i\left(\sqrt{g}g^{ij}\partial_j\Psi\right),
    \end{equation*}
being $g=det(g_{ij})$ where $(g_{ij})$ are the components of the metric tensor associated to (\ref{m2}).

The Schr\"odinger equation (\ref{SE}) reads as
\begin{equation}\label{SE1}
    \left[\frac{\partial^2}{\partial r^2}+\frac{2}r\frac{\partial}{\partial r}+\frac 1{r^2}\left(\frac{\partial^2}{\partial \theta^2}+\frac 1{k^2 \sin(\theta)^2}
    \frac{\partial^2}{\partial \phi^2}+\cot(\theta)\frac{\partial}{\partial \theta}\right)+2(h-V)\right]\Psi=0.
\end{equation}
We now put $V=-\dfrac 1r$, the Kepler-Coulomb potential, and assume the separation ansatz
\begin{equation*}
    \Psi=R(r)\Theta(\theta)\Phi(\phi).
\end{equation*}
Hence, we can obtain directly the separated equations by proceeding in the usual way:
\begin{equation}\label{R}
    R^{\prime \prime}+\frac 2r R'+\left(-\frac\lambda{r^2}+2(h-V)\right)R=0,
\end{equation}
\begin{equation}\label{T}
    \Theta^{\prime \prime}+\cot(\theta) \Theta'+\left(\lambda-\frac \mu{\sin(\theta)^2}\right)\Theta=0,
\end{equation}
\begin{equation}\label{P}
    \Phi^{\prime \prime}+k^2\mu \Phi=0.
\end{equation}
We make now the same requirements about the solutions made usually for the case $k^2=1$, following in particular \cite{Co}.

The solutions of (\ref{P}) are well defined and periodic if and only if $k^2\mu=m^2$, $m\in \mathbb Z$, then, $\mu=\dfrac {m^2}{k^2}$. It is well known that the solutions of (\ref{T}) and (\ref{P}) for $k^2=1$ are the spherical  harmonics
\begin{equation*}
  Y^m_l(\theta,\phi),
\end{equation*}
where $\lambda=l(l+1)$, are the eigenfunctions of the Laplace-Beltrami operator on $\mathbb S^2$ and $m=-l,-l+1,\ldots,0, 1, \ldots, l$, $l\in \mathbb Z$. These solutions are regular in the domain of the operator and they are physically consistent.
For the general $k$, following the same reasoning as for $k^2=1$ \cite{Co}, the (\ref{T}) and (\ref{P}) become
\begin{equation}\label{T1}
    \Theta^{\prime \prime}+\cot(\theta) \Theta'+\left(\lambda-\frac {m^2}{k^2\sin(\theta)^2}\right)\Theta=0,
\end{equation}
\begin{equation}\label{P1}
    \Phi^{\prime \prime}+m^2 \Phi=0.
\end{equation}
Therefore, in order to have regular solutions, $\dfrac {m^2}{k^2}$ must be an integer. It follows that regular solutions are obtained if and only if $k$ is a rational number. Then, the spherical harmonics in the general case can be written as
\begin{equation*}
    \tilde Y^m_l=e^{im\phi}\sin(\theta)^{\left| \frac mk\right|}\frac{d^{\left| \frac mk\right|}P_l(\cos(\theta))}{d(\cos(\theta))^{\left| \frac mk\right|}},
\end{equation*}
where
\begin{equation*}
    P_l(x)=\frac 1{2^ll!}\frac{d^l}{dx^l}(x^2-1)^l,
\end{equation*}
are the homogeneous Legendre polynomials of degree $l\geq 0$. 

In this case, the energy levels appears to be maximally degenerated, since they are parametrized by $m$ taking all integer values from $-l$ and $l$ \cite{MI}. 

We see that for $k^2\neq 1$ the parameters determining the functions $\Theta$ are different from those determining the functions $\Phi$. More importantly, we have now that the choice of $m\in \mathbb Z$ is constrained by the condition $-l\leq\dfrac mk\leq l$, $\dfrac mk\in \mathbb Z$. For example, if $k=3$, then $m$ must be integer multiple of $3$ and 
\begin{eqnarray*}
    l=0, \quad m=0, \\
    l=1, \quad m=-3,0,3,\\
    l=2, \quad m=-6,-3,0,3,6,
\end{eqnarray*}
 so on. In this case, to every eigenvalue $\lambda=l(l+1)$ corresponds again a set of $2l+1$ eigenfunctions $\tilde Y^m_l$.

Instead, the case of non integer $k$, for example $k=\dfrac 23$, gives for any $l$ 
\begin{eqnarray*}
    l=0, \quad m=0, \\
    l=1 \quad m=0,\\
    l=2, \quad m=0,\\
    l=3, \quad m=-2,0,2,\\
    l=4, \quad m=-2,0,2,\\
    l=5, \quad m=-2,0,2,\\
    l=6, \quad m=-4,-2,0,2,4,
\end{eqnarray*}
and so on. For $l>0$ the corresponding independent eigenfunctions are consequently less than $2l+1$.

Nothing changes for the functions $R$, and the consequent quantization of the energy remains the same as for $k^2=1$.

The degeneration of the energy levels is now less than maximal, meaning that some of the symmetry operators existing for $k$ integer can be lost when $k$ is not an integer. This is in accord with the results exposed in \cite{CR} for the classical Kepler system.

We recall that only rational values of $k$ allow integrability and superintegrability of a class of systems on Riemannian coverings, including the Kepler-Coulomb system \cite{CR}. 

Other considerations about degeneracy of the energy levels of the Kepler system in the case $k<1$, in dimension two, can be found in \cite{ AW}.

The superintegrability of the classical Kepler-Coulomb system on Riemannian coverings for rational values of $k$ have been proved in \cite{CR}. For the quantum case, the proof given in \cite{MPW} makes use of symmetry operators not globally defined, because of the appearance there of trigonometric functions of $k\phi$. It is expected that a globally defined set of five independent symmetry operators for the quantum case exists, proving the quantum superintegrability of the system.


\section{Klein-Gordon equation on Schwarzschild space-time}\label{s:schw}

Differently from the previous case, the Klein-Gordon equation on Schwarzschild space-time, as well as the corresponding equation of geodesics,  is not superintegrable, but only quadratically integrable, admitting separation of variables in the standard coordinates \cite{Ch,Be91}. We can see, however, as the same deficit of degeneracy seen for the Kepler-Coulomb case appears also here.

In order to see how the interpretation as covering is effective also in pseudo-Riemannian manifolds, we examine the case of the covering of a Lorentzian  four-dimensional manifold, the Schwarzschild space-time, whose metric shows a rotational symmetry. The approach is the same as in the Euclidean three-space.  
The Schwarzschild metric pulled back on a Riemannian covering can be written as \cite{Ch}
\begin{equation*}
    ds^2=(1-\frac{2M}{r})dt^2-(1-\frac{2M}{r})^{-1}dr^2
-r^2\left( d\theta^2+k^2 \sin(\theta)^2 d\phi^2\right),
\end{equation*}
where $(r,\theta,\phi)$ are spherical coordinates in the hyperplanes orthogonal to $\partial_t$ and $M$ is the mass of the black hole.

Here, the parameter $k$ may be interpreted in association with a cosmic string \cite{App,DO}.

We focus here on the the Klein-Gordon equation on Schwarzschild background. The Klein-Gordon equation becomes in this case


\begin{equation*}\label{KG}
\left[\frac{1}{1-\frac{2M}{r}}\frac {\partial^2}{\partial t^2}-\frac 1 {r^2}\frac{\partial}{\partial r}\left(r^2 (1-\frac{2M}{r})\frac {\partial}{\partial r}\right)
-\frac 1{r^2} \left(\frac{\partial^2}{\partial \theta^2}+\frac 1{k^2 \sin(\theta)^2}
    \frac{\partial^2}{\partial \phi^2}+\cot(\theta)\frac{\partial}{\partial \theta}\right)+m_P^2\right]\Psi=0,
\end{equation*}
where $m_P$ is the mass of the particle.


The equation, after assuming

\begin{equation*}
  \Psi=T(t)R(r)\Theta(\theta)\Phi(\phi),  
\end{equation*}

can be  separated in these coordinates as follows

\begin{eqnarray*}
    T^{\prime \prime}-c_2T=0\\
    \Delta^2R^{\prime \prime}+2(r-M)\frac \Delta{r^2}R^\prime+\left(\frac \Delta{r^2}c_1-\Delta m_P^2-c_2\right)R=0\\
    \Theta^{\prime \prime}+\cot(\theta)\Theta^\prime-\left( c_1+\frac {c_3}{\sin(\theta)^2}\right) \Theta=0\\
    \Phi^{\prime \prime}+k^2c_3\Phi=0,
\end{eqnarray*}
with $\Delta=1-2M/r$.
A comparison with (\ref{SE1}) shows that separation of variables leads also in this case to equations  (\ref{T1}) and (\ref{P1}), so that the same observations about the Kepler-Coulomb case  hold also for the Klein-Gordon equation on Schwarzschild background. The correspondence between the two sets of constants of motion is as follows
\begin{equation*}
    c_3=\mu,\quad c_1=-\lambda, \quad m_P^2=2h.
\end{equation*}

Several space-times show the same behaviour. 
For example, the metric \cite{DO}
\begin{equation*}
    ds^2=dt^2-b^{-2}dr^2-r^2\left(d\theta^2+k^2\sin(\theta)^2d\phi^2\right),
\end{equation*}
where $b$ is a constant and   $0<k\leq 1$, describes a space-time with a cosmic string of parameter $k$. The same modified spherical harmonics of (\ref{T1}), (\ref{P1}) are computed in a slightly different form as solutions of the Klein-Gordon equation (but any analysis of the  degeneration of energy levels is there omitted). Other examples can be found in \cite{Vil}.

\section{Conclusions}
In this note, we studied through some example the effect of considering a  Riemannian covering as a background for a quantum system. As in the classical case, we find the possibility of the loss of symmetry operators, corresponding to constants of the motions classical mechanics. These peculiar features appear in quantum system as the degree of degeneracy of the energy levels. Indeed, we have shown that the number of quantum numbers decrease for certain rational values of the covering parameter $k$, so that maximal degeneration of the energy levels of the Kepler-Coulomb system in $\mathbb E^3$ and of the Klein-Gordon equation on Schwarzschild background, become less than maximal on certain Riemannian coverings. As shown in \cite{CR}, in the classical case this corresponds to the loss of certain first integrals, in particular the quadratic ones, that may be replaced by higher order ones. We expect a similar behaviour from the symmetry operators of the quantum systems. A deeper analysis is necessary, and we leave it for future works.

\end{document}